\documentclass[sigconf]{acmart}

\usepackage{multicol}
\usepackage{multirow}
\usepackage{stfloats}
\AtBeginDocument{%
  \providecommand\BibTeX{{%
    \normalfont B\kern-0.5em{\scshape i\kern-0.25em b}\kern-0.8em\TeX}}}

\copyrightyear{2021}
\acmYear{2021}
\setcopyright{acmlicensed}\acmConference[CHI '21 Extended Abstracts]{CHI
Conference on Human Factors in Computing Systems Extended Abstracts}{May
8--13, 2021}{Yokohama, Japan}
\acmBooktitle{CHI Conference on Human Factors in Computing Systems
Extended Abstracts (CHI '21 Extended Abstracts), May 8--13, 2021, Yokohama,
Japan}
\acmPrice{15.00}
\acmDOI{10.1145/3411763.3451610}
\acmISBN{978-1-4503-8095-9/21/05}
\begin{document}


\title{Mitigating the Effects of Reading Interruptions by Providing Reviews and Previews}


\author{Namrata Srivastava}
\orcid{0000-0003-4194-318X}
\affiliation{%
  \institution{University of Melbourne, and Monash University}
  \city{Melbourne}
  \state{VIC}
  \country{Australia}
}
\authornote{This work was conducted as part of the summer internship project with Adobe Research.}
\email{srivastavan@student.unimelb.edu.au}

\author{Rajiv Jain}
\affiliation{%
  \institution{Adobe Research}
  \city{College Park}
  \state{Maryland}
  \country{United States}}
\email{rajijain@adobe.com}

\author{Jennifer Healey}
\affiliation{%
  \institution{Adobe Research}
  \city{San Jose}
  \state{California}
  \country{United States}}
\email{jehealey@adobe.com}

\author{Zoya Bylinskii}
\affiliation{%
  \institution{Adobe Research}
  \city{Cambridge}
  \state{Massachusetts}
  \country{United States}}
\email{bylinski@adobe.com}

\author{Tilman Dingler}
\affiliation{%
  \institution{University of Melbourne}
  \city{Melbourne}
  \state{VIC}
  \country{Australia}}
\email{tilman.dingler@unimelb.edu.au}

\renewcommand{\shortauthors}{Srivastava et al.}

\begin{abstract}
As reading on mobile devices is becoming more ubiquitous, content is consumed in shorter intervals and is punctuated by frequent interruptions. In this work, we explore the best way to mitigate the effects of reading interruptions on longer text passages. Our hypothesis is that short summaries of either previously read content (reviews) or upcoming content (previews) will help the reader re-engage with the reading task. Our target use case is for students who study using electronic textbooks and who are frequently mobile. We present a series of pilot studies that examine the benefits of different types of summaries and their locations, with respect to variations in text content and participant cohorts. We find that users prefer reviews after an interruption, but that previews shown after interruptions have a larger positive influence on comprehension. Our work is a first step towards smart reading applications that proactively provide text summaries to mitigate interruptions on the go.
\end{abstract}

\ccsdesc[500]{Human-centered computing~Empirical studies in HCI}
\ccsdesc[500]{Human-centered computing~Empirical studies in ubiquitous and mobile computing}
\ccsdesc[500]{Human-centered computing~Mobile phones}
\ccsdesc[500]{Human-centered computing~E-book readers}

\keywords{reading interfaces, interruptions, mobile reading}

\begin{teaserfigure}
    \centering
   \includegraphics[width=0.8\textwidth]{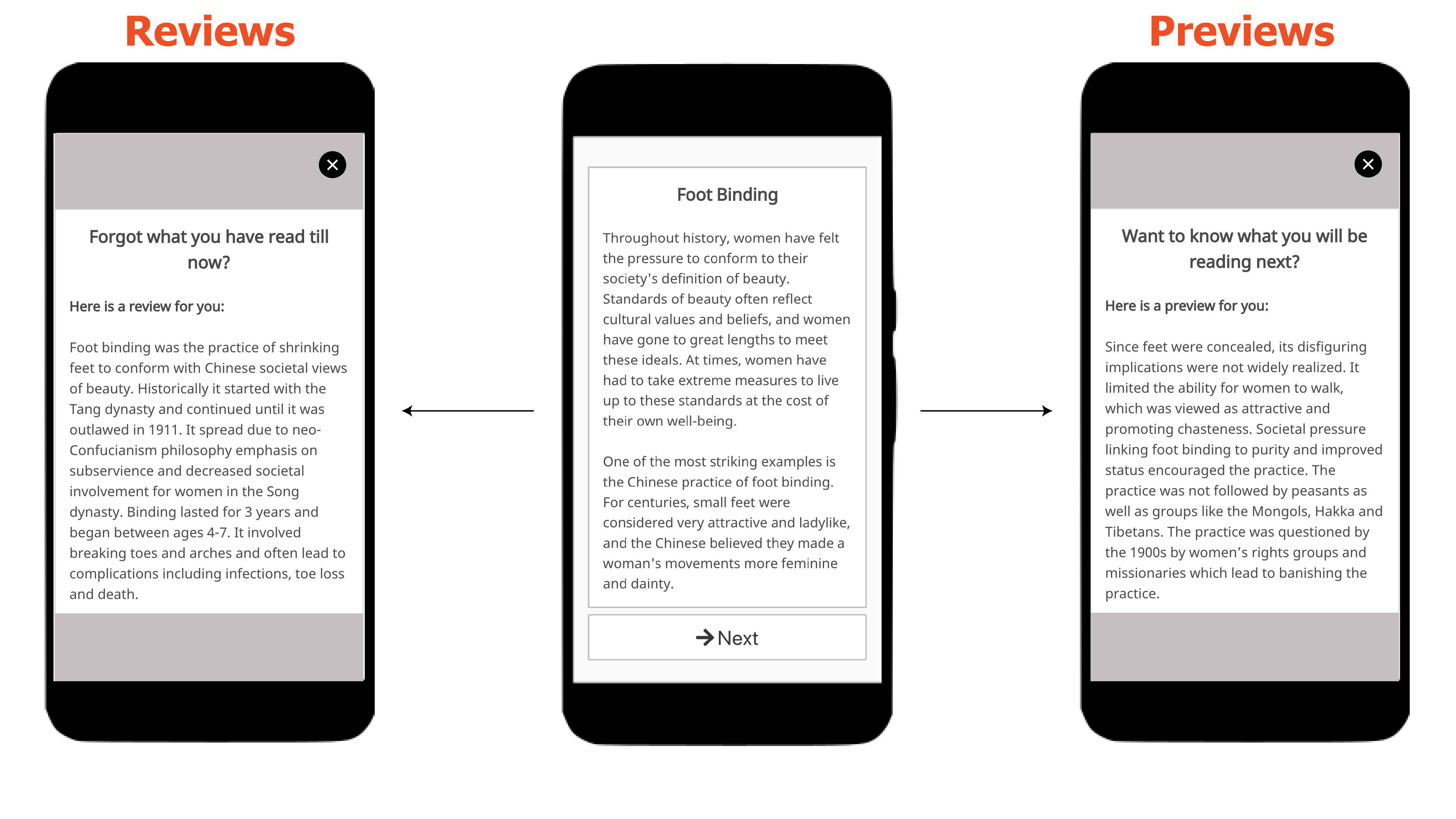}
  \caption{An example of a review and a preview used in our study for mitigating the effects of reading interruption.}
  \Description{Example of a review and a preview, where a review is a summary of the past-content, and a preview is a summary of the near future content.}
  \label{fig:review_preview}
\end{teaserfigure}

\maketitle

\section{Introduction}
Mitigating reading interruptions is a critical need for today's mobile and time-fluid lifestyles. In the past, it was generally assumed that more serious reading would take place in a quiet study environment over long periods of time, and that mobile devices would only be used for shorter, less complex tasks. In reality, despite the screen size limitations of mobile devices, many people, students included, use them for reading longer and more complex documents, papers, and textbooks~\cite{boruff2014mobile}. Students often turn to e-books as cost-effective and lightweight alternatives to paper textbooks. Mobile devices allow easy access to content and provide the ability to take and share both text and audio notes.  The mobile platform empowers people to read wherever and whenever they want: on busses and trains, at cafes, and whenever there is a few minutes to spare. A primary drawback of this reading style, however, is the likelihood of frequent interruptions, and while interruption recovery is often trivial for simple tasks, it is more difficult when the task is complex~\cite{Speier2003TheEO}.  

In readability research, the majority of work has focused on short passages, following initial benchmarks on paragraph-length reading~\cite{KintschvanDijk78}. In interruption research, the primary focus has been on studying the effects of short secondary interruptions on a continuous primary reading task, such as the appearance of a notification or text on top of the digital document. In contrast, we consider the scenario where a person is engaged in a longer, more complex reading task and needs to task-switch to a new primary task - for example, putting away the phone in order to board a bus. We wish to mitigate the effects of such major interruptions on a complex reading task, and enable the reader to quickly re-engage with the document while maintaining their comprehension. Unlike other research in smart phone and wearable computer usability research, we do not consider the user's preferences for managing the interruptions themselves~\cite{Nilsson05}, as we assume the interruptions are out of the user's control, e.g., the bus arrives and it is time to board.

In this work, we considered more complex passages between 800-1400 words. We employed fiction and non-fiction passages at eighth and twelfth grade reading levels. We created a mobile phone application that allowed participants to read passages on their own phones.  Passages were divided into several ``pages'', and an interruption was introduced mid-reading, requiring the participant to perform math, memory recall, and game tasks.  Participants were asked to complete a 10-question quiz at the end of reading to measure comprehension. Our hypothesis was that a summary of the material provided either before or after the interruption could help users re-engage with the material either more quickly or with better comprehension. We considered the efficacy of four summary presentations: a \textbf{review} summary of the first half of the passage right before the interruption, the same review presented right after the interruption, as well as a \textbf{preview} summary of the second half of the passage presented either right before or right after the interruption. It was our assumption that ``major'' interruptions, such as the need to board a bus or answer a phone call would allow at least a few seconds ``warning'' time in which a reader could glance at a summary in advance. We tested each of these conditions using different cohorts and different types of passages in three consecutive pilot studies. This study presents initial findings towards designing smart reading applications to help mitigate the effects of interruptions by providing text summaries.

\section{Related Work}
Our work is rooted in related research on mobile reading, interruptions, and task resumption. Our particular focus is on longer, more complex text passages. As text content is increasingly consumed on electronic displays, our reading habits have significantly changed. Due to their ubiquity, mobile devices are increasingly being used for reading, and while users can now read anytime and anywhere, Liu's studies~\cite{liu2005reading} show that reading activities have become rather brief and characterized by skimming behaviour, as opposed to in-depth reading sessions.

Interruptions and typical mobile behaviors pose a significant hurdle for  in-depth reading, yet interruptions as users navigate their physical environments are largely unavoidable. 
Leiva and colleagues~\cite{Leiva2012MobileInterruptions} define two types of interactions prevalent during mobile phone usage: 1) intended interruptions, \textit{i.e.}, a user-controlled switching between applications, and 2) unintended interruptions triggered by system functions or incoming messages and calls. The type of interruption generally determines its duration and resulting overhead before task resumption. Leiva et al. suggest supporting users in regaining the context of the deferred application. 

There are generally three approaches available to address and mitigate the effects of interruptions:
1) (Re-)scheduling interruptions, such as delaying notifications~\cite{Iqbal2005Interruptions}, 
2) Preparing users for interruptions by, for example, providing contextual cues for task resumption~\cite{hodgetts2006contextual}, and
3) Supporting task resumption as users return to their task. For instance, Kern et al.~\cite{Kern2010AttentionSwitching} used eye-tracking
to help users switch and resume tasks, by highlighting the last gaze position with visual placeholders. 
Mariakakis et al.~\cite{Mariakakis2015SwitchBack} designed \textit{SwitchBack}, a mobile app that uses the front-facing camera to track reading progress and highlight where the reader left off prior to an interruption.
A comprehensive design space for mobile task resumption has been described by Schneegass and Draxler~\cite{schneegass2021designing}.

Dingler et al.~\cite{Dingler17Priming} proposed the use of text summaries as priming cues to help readers navigate text more effectively. \textit{Priming} is a way of facilitating the cognitive processing of a stimulus through prior exposure to concepts related to the target stimulus~\cite{tulving1982priming}. It has been shown to help with memory encoding and retrieval~\cite{ratcliff1988retrieval} as well as facilitating text comprehension~\cite{Angerbauer2015Priming}.
Summaries prior to and after content consumption generally help comprehension and memory~\cite{baddeley2009memory}. For comprehending text, readers need to be able to recall text content as well as understand the underlying concepts~\cite{Morineau2005}. 
By mentally processing the new information, readers connect it to their prior knowledge~\cite{Ahmadi2010}. Bransford and Johnson~\cite{bransford1972contextual} argue that relevant contextual knowledge aids, and even acts, as a prerequisite for text comprehension.  

In our work, we are interested in mitigating the cognitive overhead caused by mobile interruptions by providing readers with text summaries before or after the interruption. While text summaries have been widely used to support readers' memory, we propose that reading previews can prepare readers for what is to come. We compare the effects of review summaries with preview summaries. We further investigate at which point such summaries are best shown, \textit{i.e.,} prior to or after an interruption.

\section{Method}
To study the effects of reviews and previews triggered before and after reading interruptions, we conducted a series of pilot studies. All studies used the same technical framework but varied in design, cohort type and content type.

\subsection{Study Design}
To evaluate how priming cues could help mitigate interruptions during mobile reading we started with the study design shown in Figure~\ref{fig:blockdiagram}, which consists of three main experiment blocks:

\begin{figure}[!h]
    \centering
    \includegraphics[width=0.9\columnwidth]{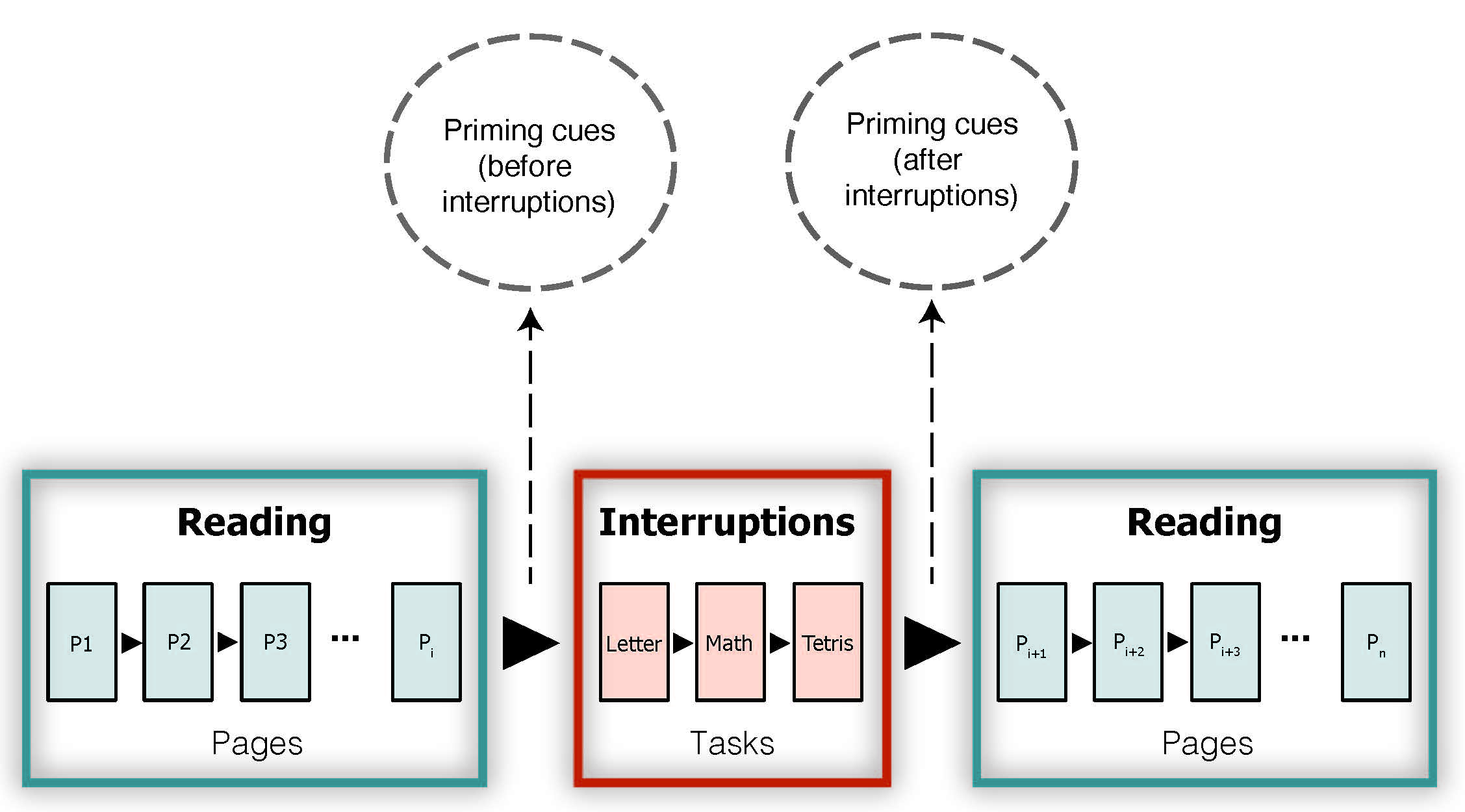}
    \caption{Our study design consists of three main experiment blocks: reading block, interruption block, and a priming cue.}
    \Description[The study design consists of three main element blocks - reading block, interruption block, and a priming cue.]{The study design consists of three main element blocks - reading block, interruption block, and a priming cue. A priming cue (preview/review) can appear either after or before the interruption}
    \label{fig:blockdiagram}
\end{figure}

\begin{enumerate}
    \item The \textbf{reading block} consists of fictional or non-fictional passages from a standard reading literature database. The reading block was split into two parts with an interruption block in between.
    
    \vspace{5pt}
    
    \item The \textbf{interruption block} consists of three cognitive tasks, as described below:
    
    \begin{itemize}
        \item In the \textbf{letter task}, participants were shown a sequence of 3 letters one at a time and were asked to remember them, before repeating them back.
        \item In the \textbf{math task}, participants were asked to perform fast addition of 1-digit numbers for 20 seconds.
        \item In the \textbf{Tetris task}, participants were asked to play a game of Tetris\footnote{https://aerolab.github.io/blockrain.js/} for 1 minute. Tetris is a block-matching game where participants score points by placing falling blocks in carefully stacked configurations.
    \end{itemize}
    
   All the interruption tasks were designed to keep the participant engaged throughout the experiment without overloading their cognitive capacities. The letter task and mental math were included as short term memory-wiping tasks, and the Tetris game was included for active distraction. 
    
    \vspace{5pt}
    
    \item Two types of \textbf{priming cues}, in the form of human-generated short summaries (i.e., 80 words) were utilized in our study for mitigating the effects of interruptions during reading:
    
    \begin{enumerate}
        \item \textbf{Reviews} consist of a short summary of the content that the participant has already read. They were designed to help the users recall and revisit the information.
        \item \textbf{Previews} consist of a short summary of the content that the participant will be reading next. They were designed to motivate the users to read further.
    \end{enumerate}
An example of a review and a preview is shown in Figure \ref{fig:review_preview}. The priming cues could appear right before or right after the interruption.
\end{enumerate}

\subsection{Implementation Details}   

\begin{figure*}[!h]
        \centering
        \includegraphics[width=0.75\textwidth]{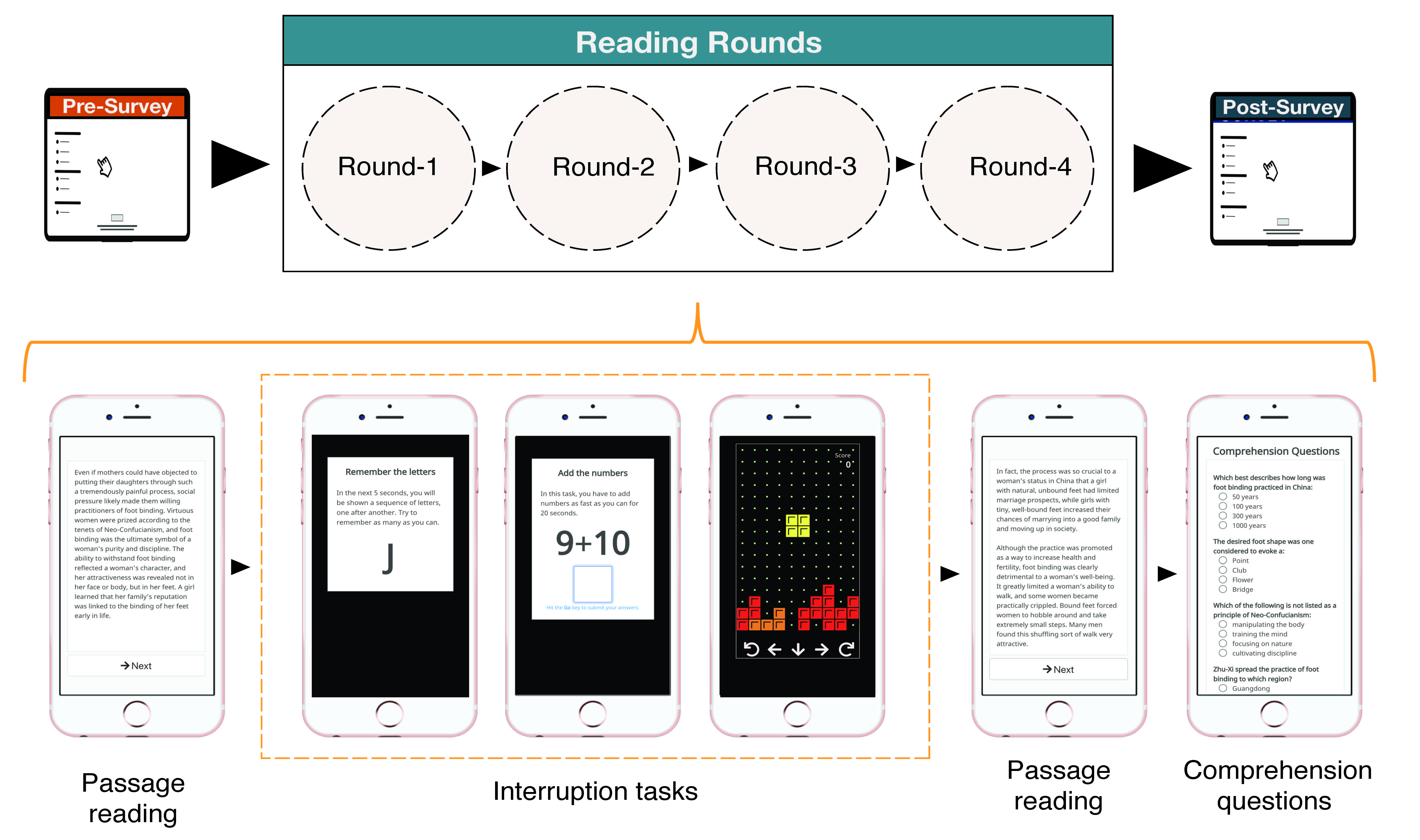}
          \caption{The basic flow of the experiment}
      \Description{Basic flow of the reading experiment consisting of multiple reading rounds. In each of the reading rounds, a passage was shown to participants with three interruption tasks in between.}
    \label{fig:basicflow}
    \end{figure*}

We developed a mobile web application using HTML5 and JavaScript, built on top of the work of Wallace et al.~\cite{wallace2020accelerating}. The application could be run on both Android and iOS devices. The basic flow of the experiment is visualized in Figure~\ref{fig:basicflow}. The experiment consists of multiple reading rounds. In each of the reading rounds, a passage was shown to participants with three interruption tasks in between. Further, after the passage, 10 multiple choice questions were added to test for reading comprehension. Reading passages were broken up into smaller pages, with a similar number of words per page (90 on average). Rather than scrolling through a longer passage, participants clicked a ``Next" button to move between the pages. A pre-survey recorded participants' demographics, and a post-survey collected subjective feedback about the previews and reviews.

All the interruption tasks were presented one after another: first the letter task, then the math task, and finally the Tetris task. To increase participant engagement, participants were shown encouraging message prompts about their performance in each task. For example, ``Perfect!'' if they correctly remembered all the letters in the letter task, or ``Better luck next time!'' if not.  Participants could not pause the interruptions or go back to the previous content during the course of the experiment.

\section{Experimental Setup}

\subsection{Pilot 1}

\begin{figure*}[!h]
    \centering
    \includegraphics[width=0.75\textwidth]{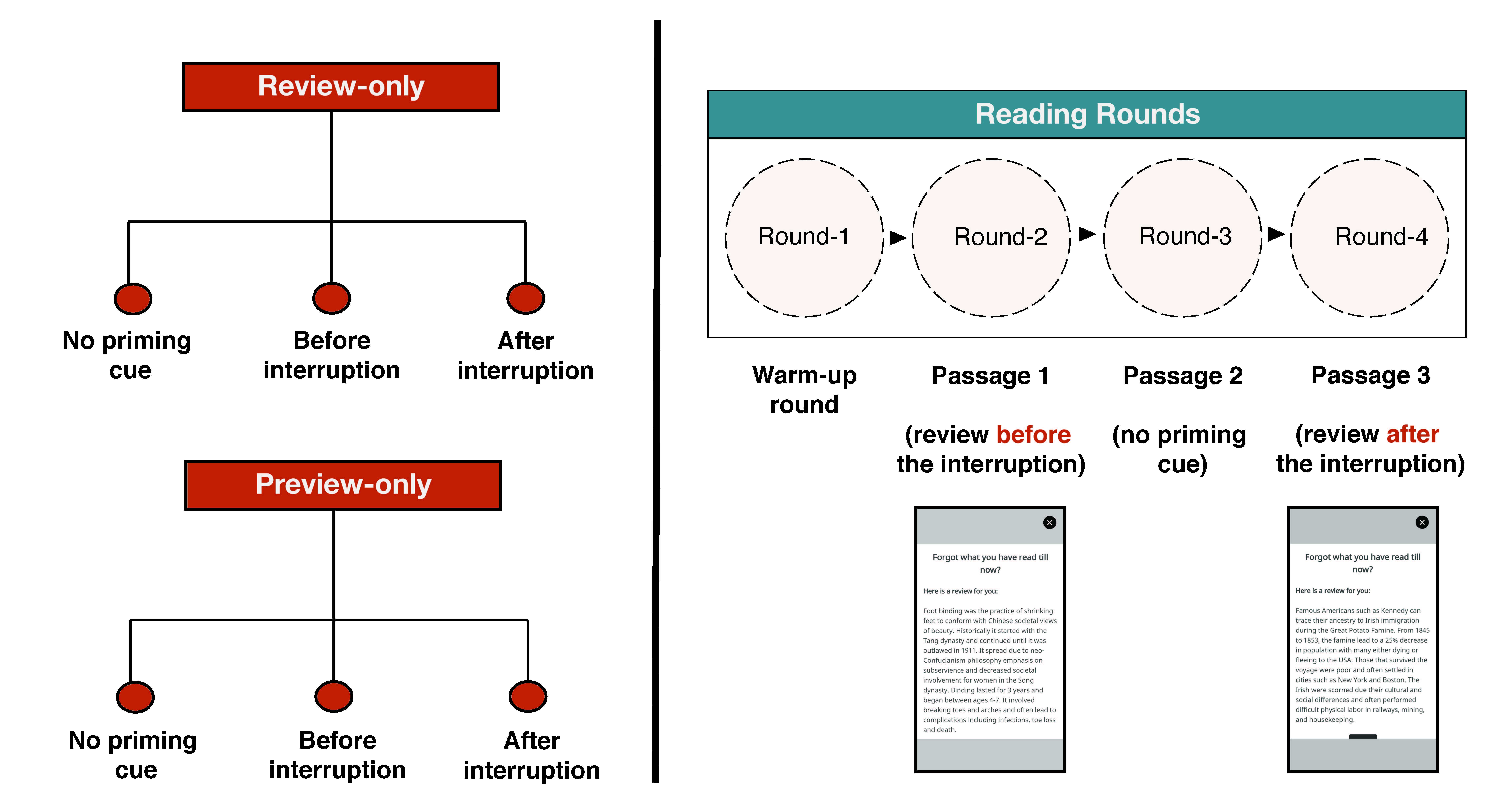}
    \caption{Left: The sub-conditions for both the preview-only and review-only reading conditions. Right: A example of the between-subjects design for the   review-only reading condition.}
    \Description{The sub-reading conditions for both the preview-only and review-only reading conditions. As shown, in both the conditions, either priming cues will be displayed before or after the interruption, or there will be no priming cues in that round.}
    \label{fig:review_example}
\end{figure*}

The first pilot study was conducted on the Amazon Mechanical Turk (MTurk) crowdsourcing platform with 20 participants who identified as native English speakers. Participants ranged in age from 28 to 56 years (average = 44), with 50\% identified as female. We followed a between-subjects study design where half of the participants were exposed to the ``preview-only'' reading condition, and the remaining half to the ``review-only'' reading condition. In both reading conditions, participants completed four rounds of reading. The first round of reading was for warm-up purposes. For the remaining three rounds, participants saw a summary (either a preview or review, depending on the participant's assigned condition) before the interruption, a summary after the interruption, and no summary at all. The ordering of these three rounds were counter-balanced across participants. 
Figure \ref{fig:review_example} contains an example.
We tested reading comprehension for shorter passages (500-600 words). Both passages and comprehension questions were chosen from ELC study zone\footnote{https://continuingstudies.uvic.ca/elc/studyzone}, a reading literature database.

\subsection{Pilot 2}

The second pilot study had a cohort that was comprised of a set of 6 university students, who were observed live while participating by a remote experimenter via video teleconference, as well as a cohort of 20 MTurk users who identified as native English speakers. Participants ranged in age from 22 to 60 years (average = 37), with 42\% identified as female. We followed the same between-subjects study design from Pilot 1. For this study, passages were selected from ReadWorks\footnote{\label{note2}https://www.readworks.org/} and were fiction, between 1000-1200 words, and at an 8th grade US reading level. These passages were chosen to understand the effect of interruptions on the reading of longer passages. 

\subsection{Pilot 3}

The third pilot study was conducted with another set of 6 university students, observed remotely, and 50 MTurk users who identified as native English speakers. Participants ranged in age from 22 to 63 years (average = 37), with 53\% identified as female. For this study, we switched to a within-subject design where each participant was tested on all of the following five conditions, in random order: preview-before, preview-after, review-before, review-after, and no-summary. Passages were again selected from ReadWorks\footnote{https://www.readworks.org/}  and were non-fiction historical literature, between 1200-1400 words, and at an 11-12th grade US reading level. These passages were chosen to understand the effect of interruptions on non-fiction passages with higher informational content and more challenging content. 

\section{Results And Discussion}

Data from the three pilot studies were analysed to evaluate how different priming cues (preview or review) and presentation locations (before or after the interruption) can mitigate interruption on comprehension of complex content. With respect to our cohorts, we observed that some MTurk workers had anomalous behavior, both reading very fast ($>$600 WPM) or taking too long to answer reading comprehension questions ($>$4 minutes). We suspected it was possible to screen save the content and then review it outside of the app. Therefore, we disqualified participants with this behavior, leaving 19 participants for the Pilot 1 study, 25 participants for the Pilot 2 study, and 48 participants for the Pilot 3 study. In addition to the statistics presented in Table \ref{tbl:reading_comprehension}, we also collected qualitative feedback about the participants' experience in the third pilot.
This allowed us to assess which priming cues were preferred, in addition to which were most effective with respect to comprehension. A brief description of all the results is provided here.

\subsection{Effect of priming cues on reading comprehension}

Here we report the effect of the five different priming conditions ( preview-after, preview-before, review-after, review-before and no-summary) on recovering from reading interruption. Statistics about participant comprehension scores and response times (calculated from 10 multiple choice questions with four options) 
 are provided in Table \ref{tbl:reading_comprehension}. 

\begin{table*}[bp]
\caption{Effect of priming cue location on reading comprehension score and response time. The bold entries correspond to the highest mean comprehension score and lowest mean response time per study. The variable \textit{n} equals the number of participants. Pilots 1 and 2 used a between-subject design, whereas Pilot 3 used a within-subject design.  }
\label{tbl:reading_comprehension}
\begin{tabular}{ccccccc}
\toprule
                        & \multicolumn{3}{c}{\textit{Comprehension Score $\mu(\sigma)$}}                                                                                                                                       & \multicolumn{3}{c}{\textit{ Response Time $\mu(\sigma)$ (Seconds)}}                                                                                                                                                 \\ \hline
                        & \textit{\begin{tabular}[c]{@{}c@{}}Pilot 1 \\ (n=19)\end{tabular}} & \textit{\begin{tabular}[c]{@{}c@{}}Pilot 2\\ (n=25)\end{tabular}} & \textit{\begin{tabular}[c]{@{}c@{}}Pilot 3\\ (n=48)\end{tabular}} & \textit{\begin{tabular}[c]{@{}c@{}}Pilot 1\\ (n=19)\end{tabular}} & \textit{\begin{tabular}[c]{@{}c@{}}Pilot 2\\ (n=25)\end{tabular}} & \textit{\begin{tabular}[c]{@{}c@{}}Pilot 3 \\ (n=48)\end{tabular}} \\ 
\midrule
\textit{Preview-after}  & \textbf{6.22 (1.48)}                                               & 8.58 (1.16)                                                       & \textbf{7.40 (1.95)}                                              & \textbf{68.9 (42.14)}                                             & \textbf{55.8 (48.8)}                                              & 92.0 (81.0)                                                        \\ 
\textit{Preview-before} & 5.89 (1.45)                                                        & 8.5 (1.09)                                                        & 6.83 (1.78)                                                       & 72.2 (32.89)                                                      & 62.2 (57.5)                                                       & \textbf{88.2 (78.6)}                                               \\ 
\textit{Review-after}   & 5.5 (0.97)                                                         & \textbf{8.62 (1.45)}                                              & 6.54 (2.27)                                                       & 86.4 (53.01)                                                      & 61.2 (52.8)                                                       & 91.0 (78.5)                                                        \\ 
\textit{Review-before}  & 5.9 (1.37)                                                         & 8.46 (1.61)                                                       & 6.85 (2.11)                                                       & 89.0 (77.69)                                                      & 61.8 (54.2)                                                       & 90.2 (79.6)                                                        \\ 
\textit{No-summary}     & 6.14 (1.16)                                                        & 8.29 (1.58)                                                       & 6.96 (1.98)                                                       & 76.85 (39.35)                                                     & 62.3 (26.51)                                                      & 91.2 (77.0)                                                        \\ 
\bottomrule
\end{tabular}
\end{table*}

We observed that during the Pilot 2 study, participants who viewed priming cues in the form of either reviews or previews scored higher than participants who didn't see any priming cues (i.e., those in the no-summary reading condition). Further, these participants also took less time to answer the comprehension questions than participants in the no-summary condition. This could indicate that priming cues may improve reading comprehension for fictional passages.

However, for the other two pilot studies (Pilot 1 and Pilot 3), participants who viewed previews after the interruption scored better than participants who saw no priming cues. However, participants in the no-summary condition performed better than participants who viewed reviews (before or after the interruption) or previews before the interruption.

Overall, our experimental results indicate that showing priming cues in the form of short summaries after the interruption improves reading comprehension and reduces response time suggesting  improved memory recall. However, the priming location and type of reading material can influence their usefulness.

\subsection{User preference for the priming cues}

In the third pilot study, we collected subjective feedback about the previews and reviews from participants by asking the following two questions:

\begin{itemize}
    \item At which location do you prefer to see previews in future reading applications?
    \item At which location do you prefer to see reviews in future reading applications?
\end{itemize}

For both these questions, participants were given the following three forced choice options: 1) After the interruption, 2) Before the interruption, and 3) No-summary. We found that for previews and reviews participants preferred the presentation after the interruption. For previews, 43.2\% participants preferred after, 29.7\% before and 27\% preferred none. For reviews, 75.7\% preferred after, 16.2\% before and 8.1\% preferred none. 

We then asked participants to express their preference for seeing either 1) only previews; 2) only reviews, 3) both, or 4) neither, and allowed an additional free text response. In contrast to the pilot results, we found that the majority of participants (41.4\%) preferred to see only reviews, 27.6\% agreed that both would be helpful for them, 20.7\% felt previews and reviews were unnecessary altogether, and the remaining 10.3\% preferred to see previews. Some of the participant comments for each of the scenarios are included below:

\begin{itemize}
    \item \textbf{Participants who preferred reviews to previews}: 
            \begin{quote}
                \textit{``Reviews were a nice summary of the content if they are displayed after the interruption. I felt that the most important points were nicely and concisely summarised in the reviews. Felt that the quiz questions were based on those summaries as well. The summaries were easy to read.	I did not like the previews. They left me confused because I haven't read that part of the story, and the detail provided did not seem enough to give me a complete idea of what is yet to come.'' -} Pid 7
            \end{quote}
    \item \textbf{Participants who preferred previews}:
    
            \begin{quote}
                \textit{``It was interesting to see what was ahead and how others summarized them.''}- Pid 37
            \end{quote}

    \item \textbf{Participants who preferred both previews and reviews}:
    
        \begin{quote}
            \textit{``Both were helpful because they both ended up exposing the reader to the same information twice, which made it easier to recall the reading material.''} - Pid 39
        \end{quote}

    \item \textbf{Participants who didn't feel the need for either}:
    
    \begin{quote}
        \textit{``I don't like anything about them.	They are repetitive and unnecessary''} - Pid 19
    \end{quote}
\end{itemize}
    
\section{Conclusion \& Future Work}
In this work, we examined the effectiveness and desirability of presenting short summaries corresponding to content previously read (reviews) and future content (previews) both before and after an interruption to a reading task. Our goal was to mitigate the impact of interruption on the continued comprehension of longer reading tasks. Results from three pilot studies indicate that summaries can improve comprehension and memory recall. Despite participant preferences for seeing reviews after an interruption, we found that showing previews after the interruption had the most consistent positive impact on improving post-interruption comprehension. The strongest result is that participants were not bothered by the summaries, and a strong majority of participants, 79.3\%, preferred to see some type of summary rather than none.
In future work, we would like to generalize the study to larger populations and conditions closer to reading in-the-wild. We also want to study the effect of additional types of priming representations beyond summaries such as lists or mind-maps. 

\begin{acks}
We would like to thank Shaun Wallace for his initial contribution in building this code.
\end{acks}

\bibliographystyle{ACM-Reference-Format}
\bibliography{PR}


\end{document}